\documentclass[letterpaper]{article}

\usepackage[T1]{fontenc}
\usepackage[utf8]{inputenc}

\usepackage{geometry}
\usepackage{setspace}

\usepackage{amsmath}
\usepackage{amssymb}
\usepackage{bm}

\usepackage[style = chem-acs, articletitle = true, doi = true]{biblatex}
\usepackage{graphicx}
\graphicspath{{./img}}   
\usepackage{subcaption}
\usepackage{float}
\usepackage{booktabs}
\usepackage{array}

\usepackage[version = 4]{mhchem}
\usepackage[hidelinks]{hyperref}
\usepackage{authblk}
\author[1]{Micha\l{} Kulczykowski}
\author[1,2]{Rafa\l{} \L{}ab\k{e}dzki}
\affil[1]{deepsense.ai}
\affil[2]{SGH Warsaw School of Economics}

\title{Self-Supervised Pretraining of Molecular Graph Encoders with LeJEPA}
\date{}

\begin{document}

\maketitle

\begin{abstract}
Self-supervised pretraining has transformed language and vision, but its value
for molecular graph neural networks remains contested; we ask whether pretraining
on a large unlabelled corpus improves downstream molecular property prediction.
To answer it we adapt LeJEPA - a predictor-free joint-embedding predictive
architecture whose embeddings are regularised by Sketched Isotropic Gaussian
Regularisation (SIGReg) - to molecular graphs, and evaluate it with two encoder
backbones (a GPS graph transformer and a Chemprop-style D-MPNN) on the Wong
et al.\ \cite{Wong2024} antibiotic-activity dataset and the \texttt{ogbg-molhiv}
benchmark under a rigorous multi-seed, bootstrap-based protocol. We find that the
benefit of pretraining is real at the level of the learned representation but does
not robustly survive finetuning. A frozen probe on the pretrained embedding far
exceeds the same architecture at random initialisation on both tasks (on
\texttt{ogbg-molhiv}, ROC-AUC $0.788$ vs $0.665$, a $+0.123$ lift reaching the
published self-supervised band), yet this edge does not convert into a downstream
finetuning gain. On the antibiotic scaffold split it is significant on a single
canonical partition ($\Delta\text{AUPRC} = +0.041$, $p = 0.010$) but vanishes when
replicated across five partitions (pooled $+0.013$, $p = 0.095$), and finetuning is
null on the random split, on \texttt{ogbg-molhiv}, and with the D-MPNN backbone.
The representational edge is nonetheless real and recoverable by other means. A
width-controlled analysis shows the embedding saturates at ${\sim}16$--$32$ effective
dimensions while the fingerprint keeps improving to $1024$ bits, and that at matched
dimensionality which representation wins depends on the split: the fingerprint leads
on validation ($0.799$ vs $0.782$ at $128$ dimensions) and trails on the shifted test
scaffolds ($0.759$ vs $0.788$), which the embedding crosses with essentially no loss
against the fingerprint's $0.040$. Truncating the embedding to its informative
subspace and concatenating it with a $1024$-bit Morgan fingerprint raises
\texttt{ogbg-molhiv} ROC-AUC from
$0.805$ to $0.832$ ($\Delta = +0.027$, $95\%$ CI $[+0.003, +0.054]$, $p = 0.014$),
while an untrained encoder put through the same pipeline gains nothing
($\Delta = -0.003$), so the pretrained embedding carries information the fingerprint
does not. We
conclude that LeJEPA pretraining yields a measurably better molecular
representation whose downstream value is realised by feature-level combination
rather than by finetuning, under which it gives at most a weak and
partition-dependent gain that a single favourable split can make look significant.
We release all pretraining code, configurations, and checkpoints.
\end{abstract}

\section{Introduction}

\subsection{Pretraining for molecular graph neural networks}

Graph neural networks (GNNs) have become the standard tool for learning over
molecular structure, treating a molecule as a graph of atoms and bonds and
producing a fixed-size embedding suitable for property prediction
\cite{gilmer2017neural}. Their accuracy, however, is bounded by the amount of
\emph{labelled} data available for the target property, and in most areas of
chemistry experimental labels are scarce and expensive, since a typical discovery
campaign measures a single endpoint on a few thousand compounds, while the
space of synthesisable, drug-like molecules numbers in the hundreds of millions
\cite{irwin2020zinc20}. This mismatch has motivated \emph{self-supervised
pretraining}, which learns general-purpose molecular representations from large
pools of \emph{unlabelled} structures and then transfers them to data-poor
downstream tasks, mirroring the now-dominant ``pretrain-then-finetune'' recipe
of language and vision.

Three families of self-supervised objectives have been explored for molecular
graphs. \emph{Contrastive} methods pull together augmented views of the same
molecule while pushing apart views of different molecules, but require carefully
designed augmentations and large numbers of negative samples. \emph{Generative}
methods, most prominently masked auto-encoding, reconstruct masked atoms or
bonds in input space \cite{hou2022graphmae}; reconstruction forces the model to
capture fine input detail that is not always useful downstream.
\emph{Predictive} methods, including the node- and graph-level attribute and
context prediction of Hu et al.\ \cite{hu2019strategies}, sit between the two.
Across these families the empirical picture is mixed. Pretraining helps on some
endpoints and is neutral or harmful on others, and a well-tuned supervised
baseline over hand-crafted molecular descriptors is often hard to beat
\cite{wang2022molecular}. This makes a careful, mechanistic study of \emph{when}
and \emph{why} a given pretraining objective transfers as valuable as a new
objective itself.

\subsection{World models and LeJEPA}

A complementary line of motivation comes from \emph{world models}, predictive
models of an environment that an agent can use to anticipate the consequences of
its actions and to reason and plan in an internal latent space, rather than by trial and error in the world itself
\cite{lecun2022path, maes2026leworldmodel}. Joint Embedding Predictive
Architectures (JEPAs) are a particularly attractive way to build such models.
Instead of modelling every aspect of an observation, a JEPA encodes observations
into a compact, low-dimensional latent space and, from the rest, predicts the
\emph{latent} representation of a held-out part - a future observation, or another
view. By predicting in representation space it focuses on the features that
are actually predictable while discarding incidental, unpredictable detail; this
distinguishes it from generative models, which must reconstruct the input in full,
including detail that is often irrelevant, and from contrastive methods, which rely
on large numbers of negative samples whose required count grows with the embedding
dimension (the curse of dimensionality)~\cite{lecun2022path}.

The central difficulty in training a JEPA is \emph{representation collapse}, whereby the
prediction objective is trivially satisfied by mapping every input to the same
constant embedding. Established JEPAs prevent this with asymmetric
machinery - a momentum (exponential-moving-average, EMA) teacher encoder,
stop-gradients, and a deliberately weakened predictor network; the graph-domain
instantiation Graph-JEPA \cite{skenderi2023graph} follows exactly this recipe,
predicting the latent codes of masked subgraphs with an EMA target encoder. The
Latent-Euclidean JEPA (LeJEPA)
\cite{balestriero2025lejepaprovablescalableselfsupervised} removes all of it,
replacing the teacher, stop-gradient and predictor with a single explicit
regulariser - Sketched Isotropic Gaussian Regularisation (SIGReg) - that drives
the embedding distribution toward an isotropic Gaussian. Collapse then becomes
impossible by construction, since a collapsed distribution is not isotropic, and
the objective reduces to a transparent sum of a prediction term and a
distributional regulariser with no asymmetric components to tune. The same SIGReg
recipe has recently been shown to train stable world models for control directly
from pixels (LeWorldModel \cite{maes2026leworldmodel}), underscoring that it is a
general, modality-spanning approach to predictive representation learning. LeJEPA
itself has so far been demonstrated on image data - natural and domain-specific
image classification, with its pretrained features additionally supporting
emergent unsupervised video object segmentation - and, to our knowledge, has not
previously been applied to molecular graphs. This work adapts it to the
molecular domain and studies, with deliberately rigorous controls, what its
pretrained representations do and do not buy on downstream chemistry tasks.

\subsection{Antibiotics and antimicrobial resistance}

Our primary application is antibiotic discovery. Antimicrobial resistance (AMR)
is among the most pressing threats to public health. Resistant infections
already cause hundreds of thousands of deaths annually and, on current trends,
threaten to render large parts of the antibiotic arsenal ineffective
\cite{o2014antimicrobial,TangAMR, Aslam2024AMR, Amann175AMR}. Yet the antibiotic pipeline has languished for
decades - the chemical scaffolds that make good antibiotics are rare, and
classical screening of large libraries is slow and costly
\cite{walsh2003will}. Machine learning has recently re-energised the field.
Stokes et al.\ \cite{stokes2020deep} used a graph model to identify halicin, a
structurally novel antibiotic, and Wong et al.\ \cite{Wong2024} screened tens of
thousands of compounds for activity against \textit{Staphylococcus aureus}
alongside human-cell cytotoxicity, releasing a benchmark of $39{,}312$ compounds
with four binary endpoints. Antibiotic activity is extremely imbalanced - only
about $1\%$ of screened compounds are active - so the task is exactly the
low-label regime in which transferable, pretrained representations should, in
principle, help most.

\subsection{A standard benchmark: \texttt{ogbg-molhiv}}

Because a single niche assay is a weak basis on which to judge a pretraining
method, we additionally evaluate on \texttt{ogbg-molhiv}, a standard molecular
property-prediction benchmark from the Open Graph Benchmark \cite{hu2020open}.
It contains ${\sim}41{,}000$ molecules labelled for inhibition of HIV
replication, with a fixed, canonical Bemis-Murcko scaffold split and ROC-AUC as
the evaluation metric. Being on-distribution for our unlabelled pretraining
corpus and equipped with many published scratch-versus-pretrained baselines, it
lets us place LeJEPA pretraining in the context of the wider graph
self-supervised-learning literature.

\section{Method}

\subsection{Overview}

We pretrain a molecular graph encoder $f_\theta$ that maps a molecule
$G = (V, E)$, with atoms as the vertices $V$ and bonds as the edges $E$, to a
$K$-dimensional embedding $\bm{z} = f_\theta(G) \in \mathbb{R}^{K}$. From each
molecule we generate several \emph{views} - lightly corrupted whole-molecule
graphs and connected subgraphs - and train the encoder so that the embeddings of
all views of a molecule are mutually predictive, while
the overall distribution of embeddings is held close to an isotropic Gaussian.
Following LeJEPA \cite{balestriero2025lejepaprovablescalableselfsupervised}, this
is achieved \emph{without} a predictor network, a momentum teacher, or
stop-gradients. The entire objective is a convex combination of a prediction loss
$\mathcal{L}_{\text{pred}}$ and a distributional regulariser
$\mathcal{L}_{\text{SIGReg}}$, weighted by a single scalar $\lambda$ (we use
$\lambda = 0.2$). We define the two terms in the following subsections and state
the full objective in Eq.~\ref{eq:total}. They play complementary roles.
$\mathcal{L}_{\text{pred}}$ makes the representation informative (views of the same
molecule agree), and $\mathcal{L}_{\text{SIGReg}}$ makes it non-degenerate (the
embedding distribution stays full-rank and well-spread), so that the objective
cannot be satisfied by collapsing all molecules to a single point.

\subsection{View generation}
\label{sec:views}

\begin{figure}[t]
  \centering
  \includegraphics[width=\linewidth]{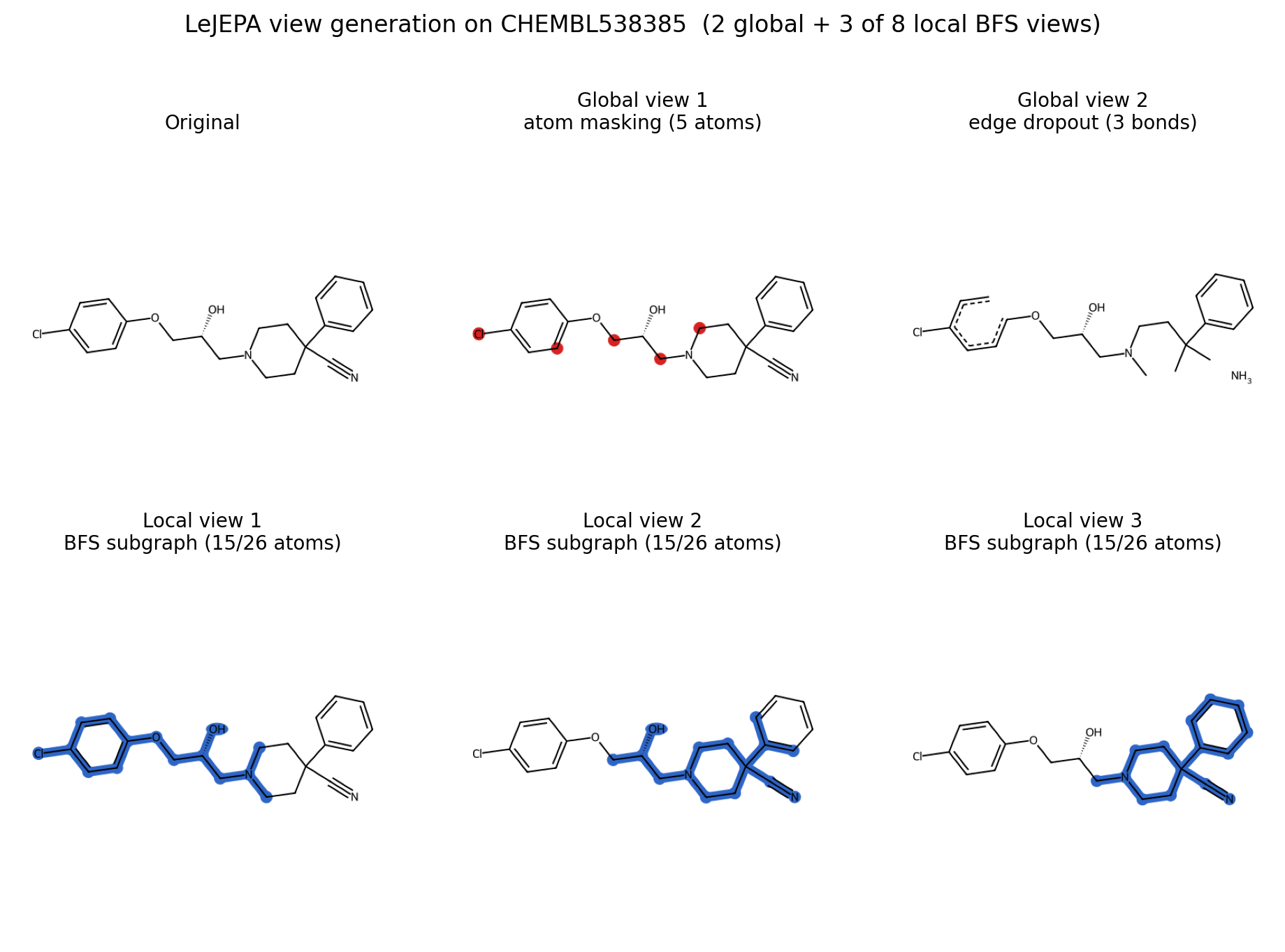}
  \caption{View generation for one molecule. Two global views (whole molecule
  with atom-feature masking and edge dropout) provide a near-whole-molecule
  reference; eight local views are connected BFS subgraphs covering ${\sim}60\%$
  of the atoms. The encoder must predict the global reference from each local
  fragment.}
  \label{fig:views}
\end{figure}

View design is the heart of the method. For each molecule we generate
$V_g = 2$ \emph{global views} and $V_l = 8$ \emph{local views} (Figure~\ref{fig:views}).

\paragraph{Global views.} A global view is the whole molecule with a light
corruption that removes a small amount of information without changing its
identity. The first global view applies \emph{atom-feature masking}, zeroing the
feature vectors of a random $15\%$ of atoms. The second applies
\emph{edge dropout}, removing a random $10\%$ of bonds from the graph. Because
both views retain essentially the entire molecule, their embeddings are nearly
identical and serve as a stable, near-whole-molecule reference.

\paragraph{Local views.} A local view is a connected subgraph grown by a
breadth-first search (BFS) from a random seed atom until it covers a target
fraction \texttt{cover\_frac} of the molecule's atoms (we use $0.6$). Local views
are the source of difficulty in the objective. Predicting the whole-molecule
reference from a $60\%$ fragment forces the encoder to learn how local
substructure determines global molecular identity.

\paragraph{Partial auxiliary conditioning.} In addition to the learned graph
features, we make a fixed vector of $217$ z-scored RDKit physico-chemical
descriptors available to the encoder. Crucially, these descriptors are attached
to the \emph{global views only}; local views receive a zero descriptor vector.
This asymmetry is deliberate. The descriptor vector is a molecule-level constant,
identical across all views of a molecule; attaching it to \emph{every} view
creates a trivial solution in which the encoder ignores structure and matches
views on the shared descriptor signal, so the prediction loss collapses to zero
without learning anything structural. Restricting descriptors to the global views
enriches the prediction \emph{target} (the global-view centroid) while leaving
the local views to be explained from structure alone, which preserves the
part-to-whole learning signal.

\subsection{Prediction loss}

Let the $V = V_g + V_l$ views of a molecule be encoded to embeddings
$\bm{z}_1, \dots, \bm{z}_V$, with the first $V_g$ being the global views. We
define the prediction target as the centroid of the \emph{global-view}
embeddings,
\begin{equation}
  \bm{\mu} \;=\; \frac{1}{V_g}\sum_{v=1}^{V_g} \bm{z}_v ,
\end{equation}
and pull every view's embedding toward it,
\begin{equation}
  \label{eq:pred}
  \mathcal{L}_{\text{pred}}
  \;=\; \frac{1}{V}\sum_{v=1}^{V} \big\lVert \bm{\mu} - \bm{z}_v \big\rVert_2^2 .
\end{equation}
There is no separate predictor network and no stop-gradient; gradients flow
through both $\bm{\mu}$ and each $\bm{z}_v$. Because the global views are
near-whole-molecule and carry the auxiliary descriptors, $\bm{\mu}$ is a stable,
informative target; the bulk of the learning signal therefore comes from the
local-view terms, which require inferring the whole-molecule centroid from a
fragment.

\subsection{SIGReg loss}

\begin{figure}[t]
  \centering
  \includegraphics[width=\linewidth]{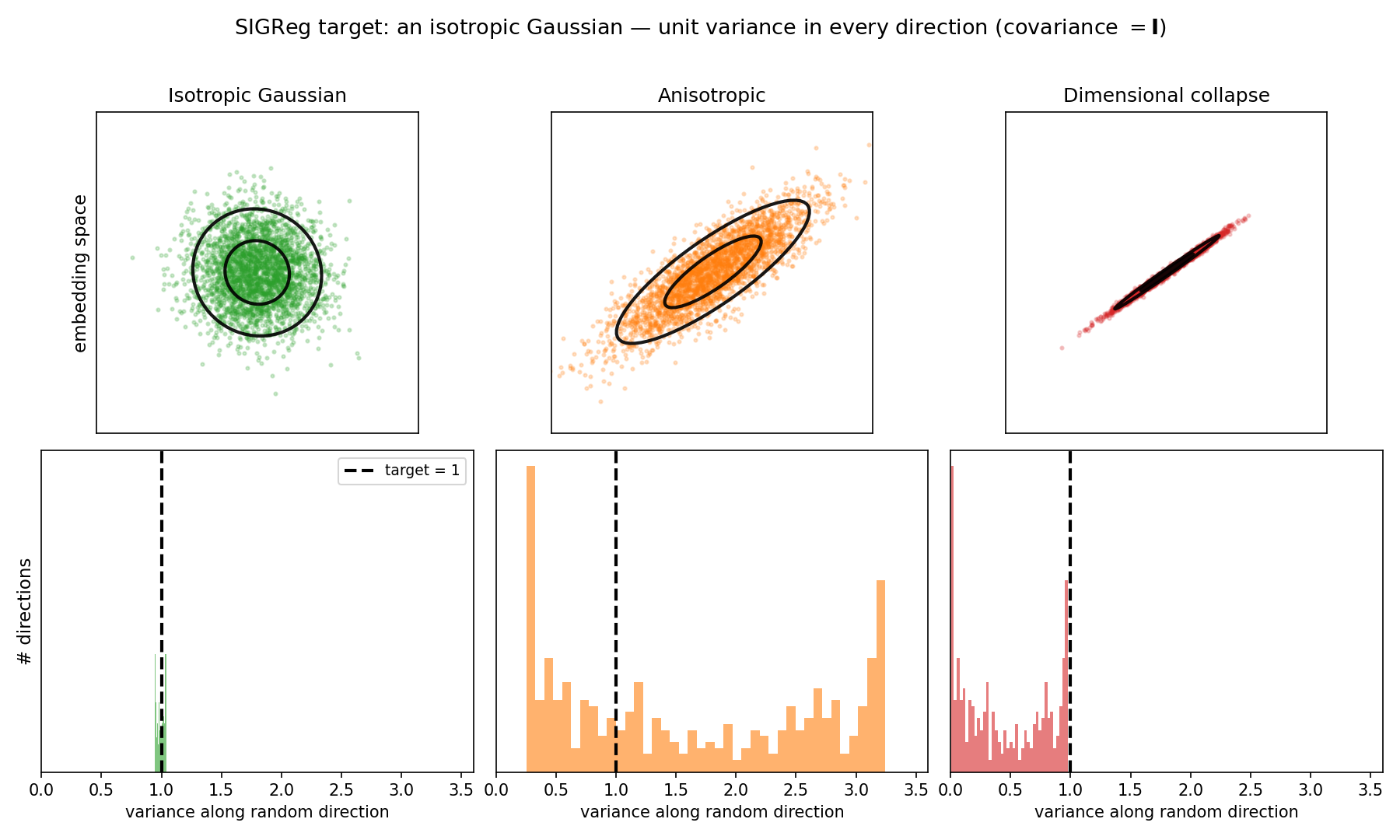}
  \caption{The SIGReg target. An isotropic Gaussian (left) has unit variance in
  every direction (covariance $=\mathbf{I}$); the failure modes SIGReg prevents
  are anisotropy (middle) and dimensional collapse (right). Bottom, the variance
  of the cloud along many random one-dimensional directions - isotropy means
  every such projection has unit variance.}
  \label{fig:isotropic}
\end{figure}

\begin{figure}[t]
  \centering
  \includegraphics[width=\linewidth]{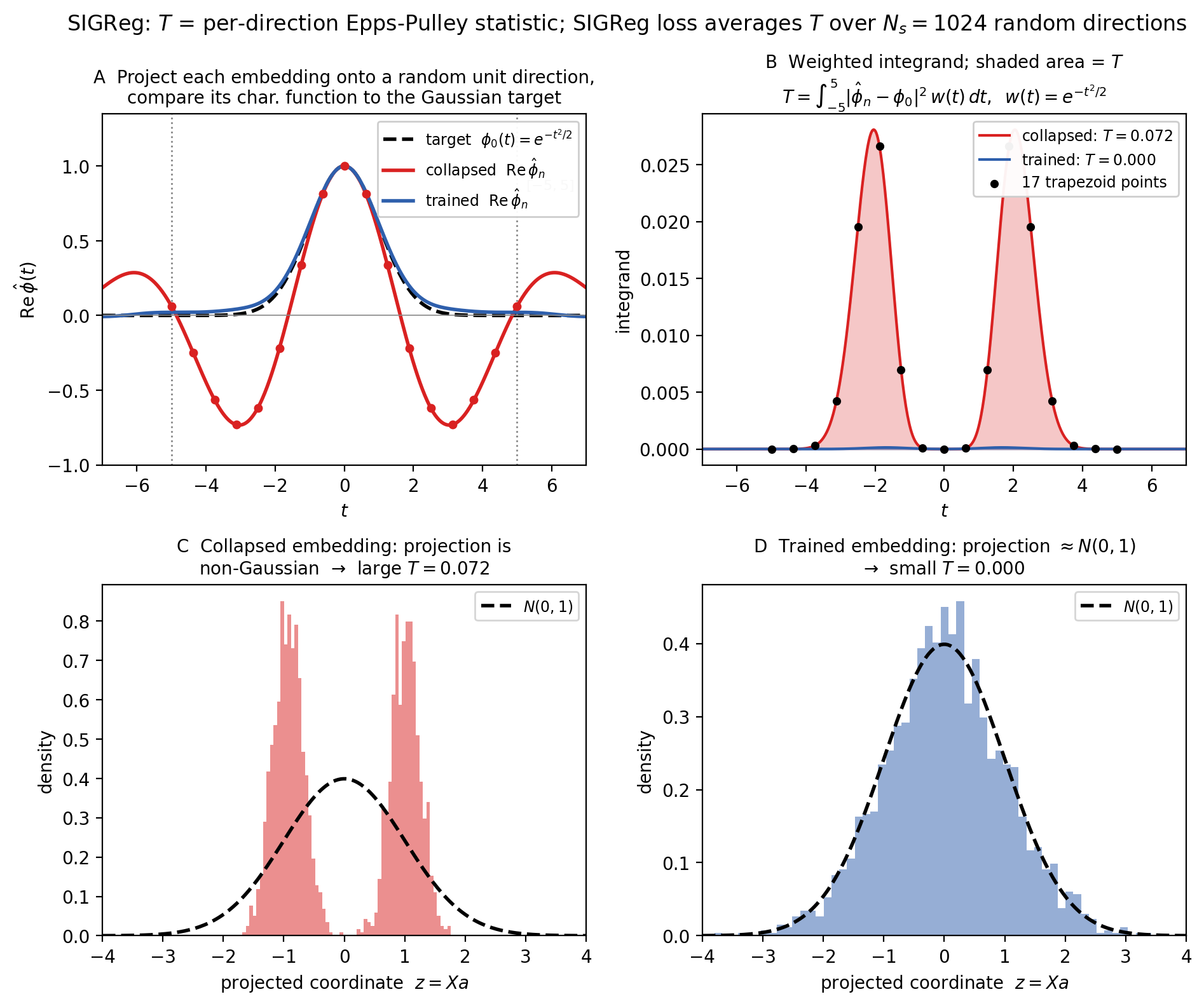}
  \caption{SIGReg in detail. Each embedding batch is projected onto random unit
  directions; per direction, the empirical characteristic function of the
  projection is compared to that of $\mathcal{N}(0,1)$ via an Epps--Pulley
  statistic $T$ (averaged over $N_s = 1024$ directions). A collapsed embedding
  yields non-Gaussian projections and large $T$; a well-trained embedding yields
  standard-normal projections and $T \to 0$.}
  \label{fig:sigreg}
\end{figure}

On its own, Eq.~\ref{eq:pred} is trivially minimised by mapping every molecule to
the same point. SIGReg \cite{balestriero2025lejepaprovablescalableselfsupervised}
prevents this by driving the batch of embeddings toward an isotropic Gaussian
$\mathcal{N}(\bm{0}, \mathbf{I})$ - a distribution that is full-rank,
unit-variance in every direction, and uncorrelated across dimensions
(Figure~\ref{fig:isotropic}). It exploits a classical fact, that a distribution is an
isotropic Gaussian if and only if \emph{every} one-dimensional projection of it
is a standard normal. SIGReg therefore ``sketches'' the embedding distribution by
projecting it onto a large number of random unit directions ($N_s = 1024$) and,
for each direction, measuring the deviation of the projected samples from
$\mathcal{N}(0,1)$ with a characteristic-function (Epps--Pulley) statistic. The
regulariser is the average of these per-direction statistics
(Figure~\ref{fig:sigreg}). We adopt the SIGReg formulation of Balestriero et al.~\cite{balestriero2025lejepaprovablescalableselfsupervised} unchanged.
Concretely, for a batch of $M$ embeddings $\{\bm{z}_n\}_{n=1}^{M}$ and a random unit direction
$\bm{u}_j \in \mathbb{S}^{K-1}$, we compare the empirical characteristic function
of the scalar projections $\bm{u}_j^\top \bm{z}_n$ with that of a standard normal,
$\phi^\ast(t) = e^{-t^2/2}$, through the univariate Epps--Pulley statistic
(Eqs.~\ref{eq:epps}--\ref{eq:sigreg} restate their definitions)
\begin{equation}
  \hat\phi_j(t) = \frac{1}{M}\sum_{n=1}^{M}
      e^{\,\mathrm{i}\,t\,\bm{u}_j^\top \bm{z}_n},
  \qquad
  T(\bm{u}_j) = \int_{-\infty}^{\infty}
      \bigl|\hat\phi_j(t) - \phi^\ast(t)\bigr|^2\, w(t)\,\mathrm{d}t ,
  \label{eq:epps}
\end{equation}
with Gaussian window $w(t) = e^{-t^2/2}$ (which gives the integrand effectively
compact support, so it is evaluated by the trapezoidal rule over $t \in [-5, 5]$).
SIGReg averages this statistic over the $N_s = 1024$ random directions,
\begin{equation}
  \mathcal{L}_{\text{SIGReg}}
  \;=\; \frac{1}{N_s}\sum_{j=1}^{N_s} T(\bm{u}_j) .
  \label{eq:sigreg}
\end{equation}
By the Cram\'er--Wold theorem, matching every one-dimensional projection to
$\mathcal{N}(0,1)$ is equivalent to matching the full joint distribution to
$\mathcal{N}(\bm{0}, \mathbf{I})$, so driving $\mathcal{L}_{\text{SIGReg}} \to 0$
makes the embedding isotropic Gaussian, which both prevents collapse and keeps the
representation well-conditioned. The statistic is computed over the global batch
under distributed training by averaging the per-direction empirical
characteristic functions across replicas. We track the
\emph{effective rank} of the embedding covariance as a health metric; it rises
toward $K$ as SIGReg takes effect, and we use no learning-rate warmup - a warmup, in our experience,
would let the prediction term collapse the representation before SIGReg engages. Putting the two terms together, the complete LeJEPA objective is
\begin{equation}
  \mathcal{L}
  \;=\; (1-\lambda)\,\mathcal{L}_{\text{pred}}
  \;+\; \lambda\,\mathcal{L}_{\text{SIGReg}} ,
  \label{eq:total}
\end{equation}
optimised end-to-end with $\lambda = 0.2$.

\subsection{Encoder architectures}

The objective in Eq.~\ref{eq:total} is agnostic to the encoder; it requires only
a map from a graph to a fixed-size embedding. We study two backbones.

\begin{figure}[p]
  \centering
  \begin{subfigure}[t]{0.49\textwidth}
    \centering
    \includegraphics[width=\linewidth]{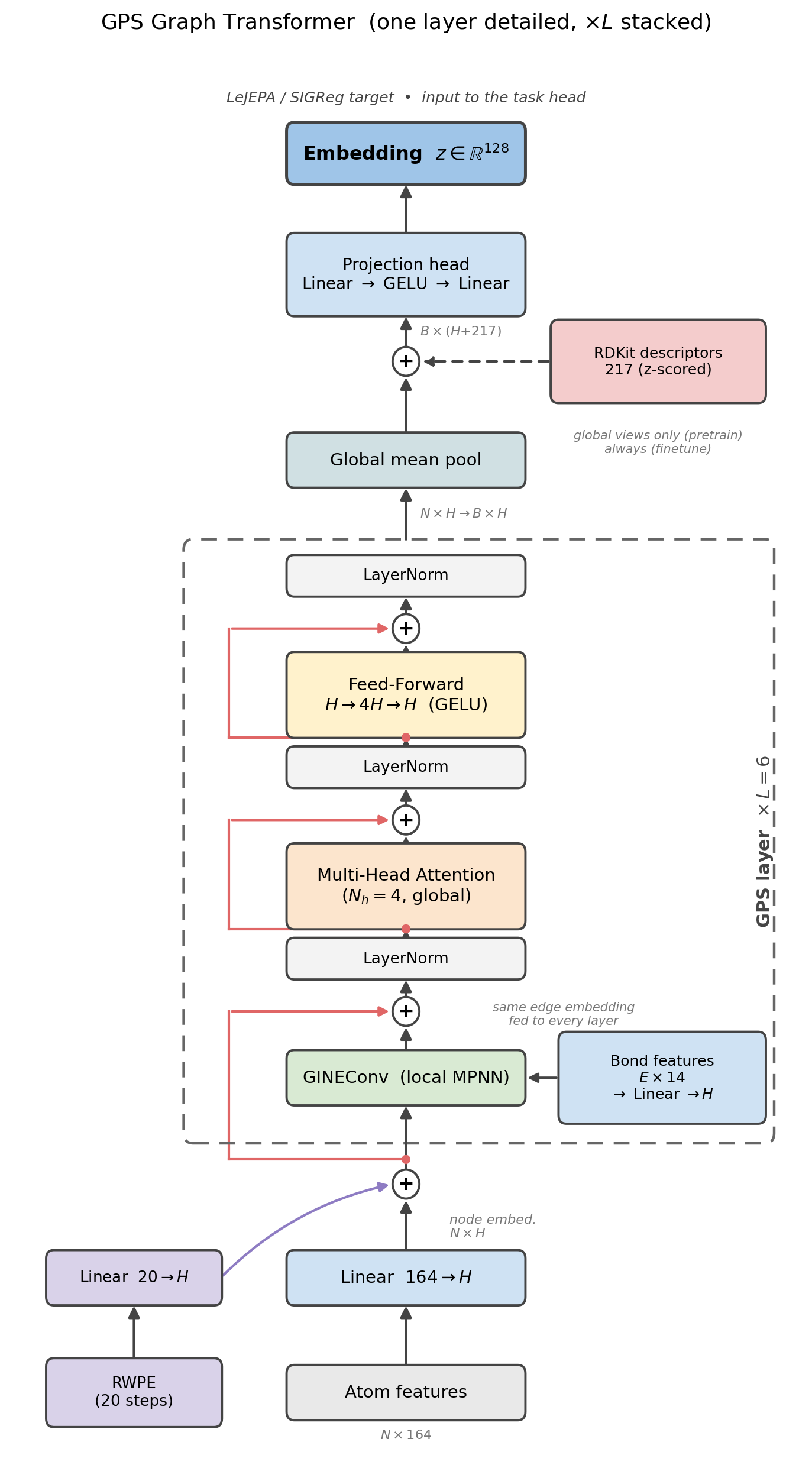}
    \caption{GPS graph transformer.}
    \label{fig:gps-a}
  \end{subfigure}
  \hfill
  \begin{subfigure}[t]{0.49\textwidth}
    \centering
    \includegraphics[width=\linewidth]{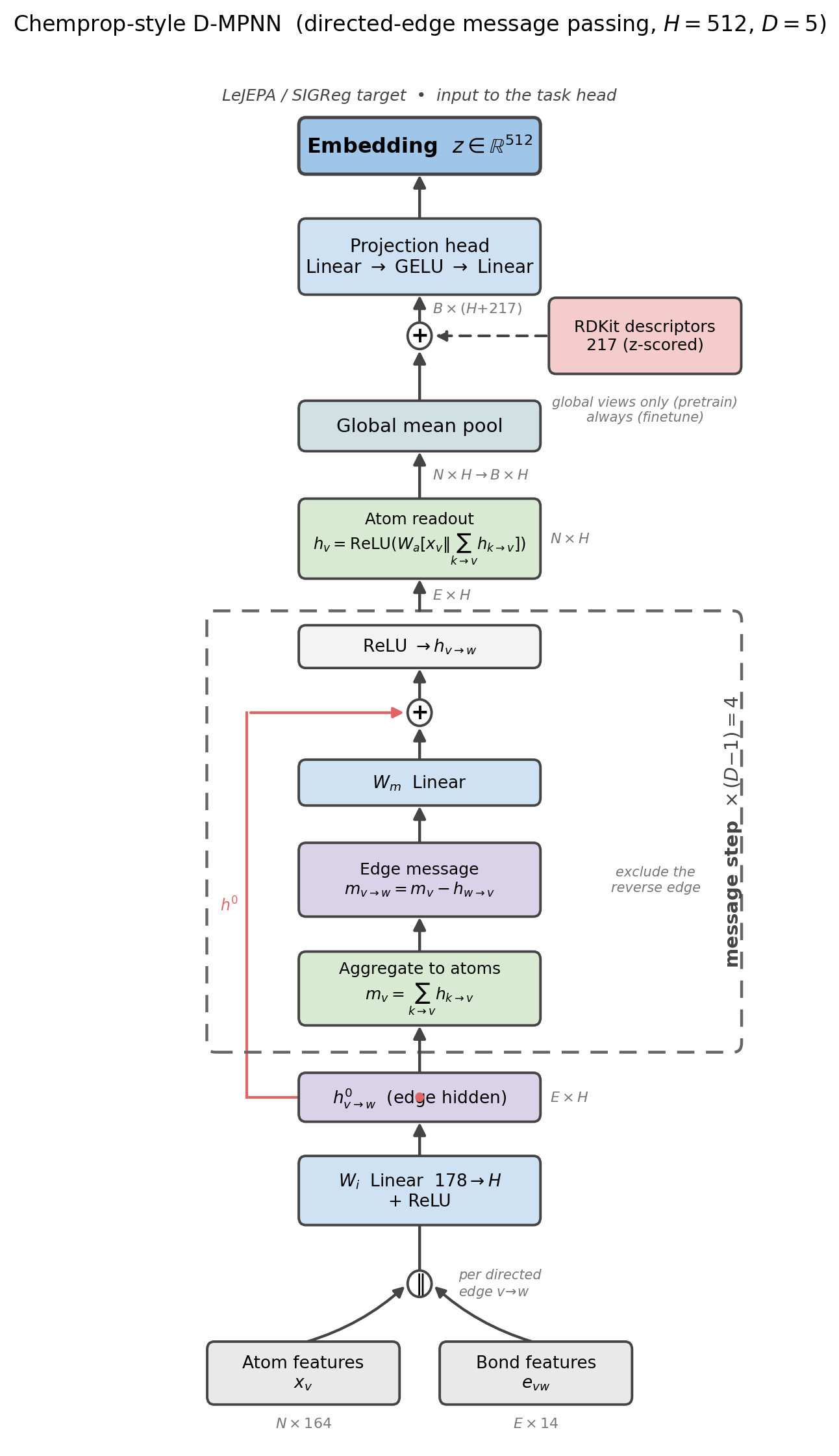}
    \caption{Chemprop-style D-MPNN.}
    \label{fig:gps-b}
  \end{subfigure}
  \caption{The two encoder backbones, sharing an identical input featurization,
  descriptor pipeline, projection head, and training objective so that only the
  message-passing core differs. \textbf{(a)} The GPS graph transformer, in which each
  layer fuses a GINE local message-passing branch with a global self-attention
  branch; random-walk positional encodings are added at the input, and node
  embeddings are pooled to the molecule embedding. \textbf{(b)} The
  Chemprop-style D-MPNN, in which messages are passed along directed bonds (each edge
  message excludes its own reverse edge) and the final atom states are pooled.
  In both, descriptors are concatenated for global views before the projection
  head.}
  \label{fig:gps}
\end{figure}

\paragraph{GPS Graph Transformer.} Our primary encoder is a GPS-style graph
transformer \cite{rampavsek2022recipe} (Figure~\ref{fig:gps}a). Each layer
combines local message passing (a GINE convolution, which incorporates bond
features) with a global multi-head self-attention block, so that every layer
mixes short-range chemistry with long-range, whole-molecule context. Atoms are
given a random-walk positional encoding (RWPE) \cite{dwivedi2022lspe}
to break symmetries that message passing alone cannot. Node embeddings are
mean-pooled to a graph embedding, optionally concatenated with the descriptor
vector for global views, and projected to the $K$-dimensional output. We use a
compact configuration (hidden dimension $128$, $6$ layers, $4$ attention heads,
$K = 128$; ${\sim}2$M parameters).

\paragraph{Chemprop-style D-MPNN.} As an architecture control we also use a
directed message-passing network (D-MPNN) in the style of Chemprop
\cite{yang2019analyzing} (Figure~\ref{fig:gps}b), reimplemented natively so that it consumes the same
graph inputs and plugs into the same objective. Messages are passed along
directed bonds (the message into an edge excludes its own reverse edge), and the
final atom states are pooled to a graph embedding. We use a larger configuration
here (hidden width $512$, depth $5$) to match the capacity at which this
architecture is competitive on supervised tasks.

\subsection{Pretraining corpus and optimisation}

Both encoders are pretrained on ${\sim}2.9$M bioactivity-curated molecules from
ChEMBL. We use this corpus rather than a larger enumerated library because it
covers substantially more of the downstream scaffolds relevant to our tasks.
For \texttt{ogbg-molhiv}, which we can quantify exactly against its fixed split,
ChEMBL contains $90.8\%$ of the training scaffolds and $44.7\%$ of the test
scaffolds, and the median nearest-neighbour Tanimoto similarity between a test
molecule and the corpus is $0.43$. This coverage matches or exceeds that of the
antibiotic task, so a null transfer result reflects the pretraining method rather
than a corpus mismatch; the roughly $55\%$ of test scaffolds absent from ChEMBL
represent a genuine but unexceptional transfer ceiling. We optimise
Eq.~\ref{eq:total} with Adam, $\lambda = 0.2$, $V_g = 2$, $V_l = 8$,
$N_s = 1024$ random projections, and no learning-rate warmup, using distributed
data-parallel training across four GPUs. All pretraining code, configurations,
and final checkpoints are released to support reproducibility.

\subsection{Qualitative effect of pretraining}

\begin{figure}[t]
  \centering
  \includegraphics[width=\linewidth]{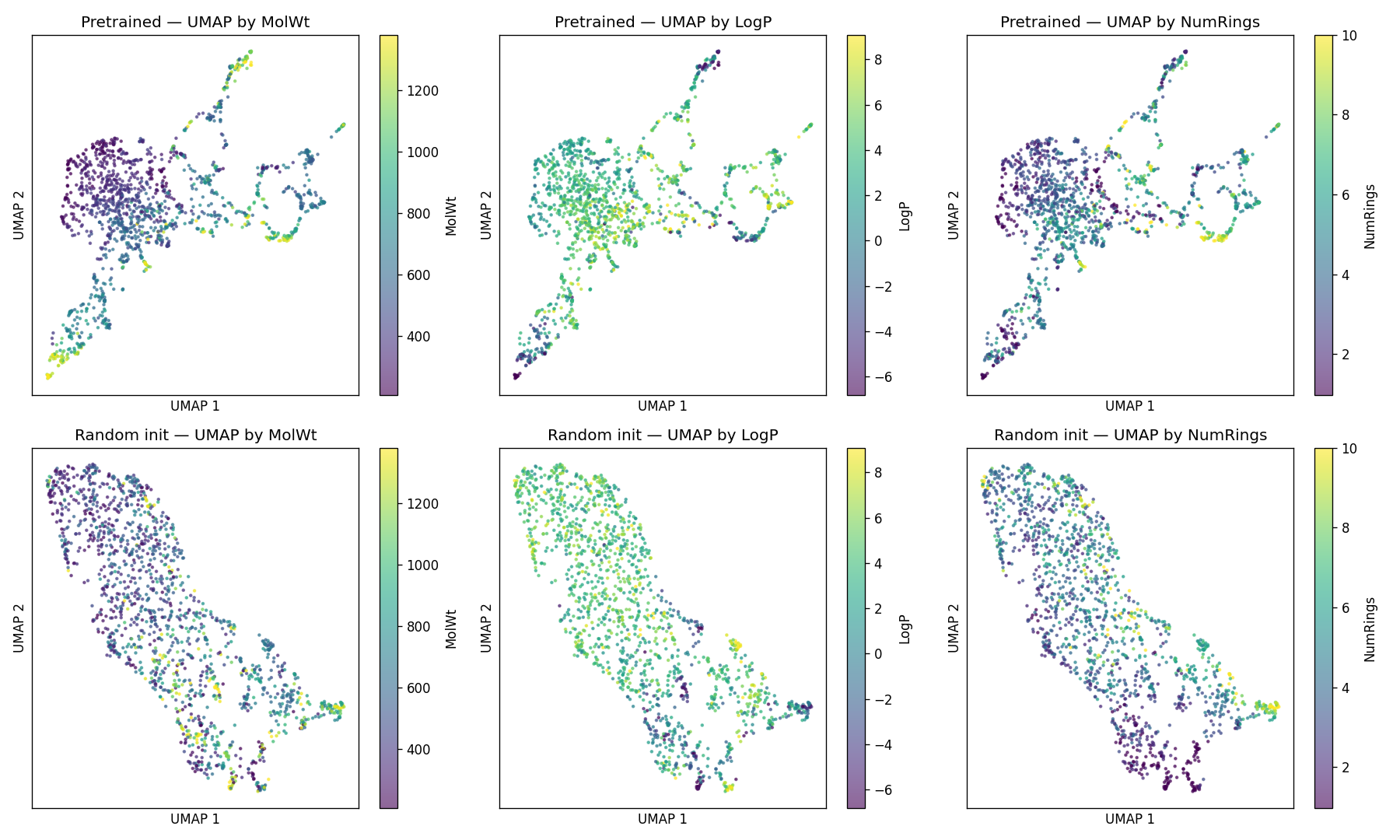}
  \caption{Two-dimensional projection of molecular embeddings, coloured by
  molecular weight, lipophilicity (LogP) and ring count, for the LeJEPA-pretrained
  encoder (top) versus a random-initialised encoder of the same architecture
  (bottom). Both use the pure graph embedding with the auxiliary descriptors
  zeroed. Pretraining produces a well-spread manifold organised by all three
  properties; the random encoder recovers only a coarse molecular-size
  (molecular-weight) axis.}
  \label{fig:embcompare}
\end{figure}

A first, qualitative check that pretraining learns chemically meaningful
structure is shown in Figure~\ref{fig:embcompare}, which projects the embeddings
of ChEMBL molecules to two dimensions for the pretrained encoder and for a
randomly-initialised encoder of the same architecture. To isolate the
\emph{learned} representation, both projections use the pure graph embedding with
the auxiliary RDKit descriptors zeroed. Otherwise the descriptors, which directly
encode these properties, would trivially organise either embedding. The
pretrained embedding forms a well-spread manifold with smooth gradients in several
physico-chemical properties (molecular weight, lipophilicity, ring count). The
random-initialisation embedding recovers only a coarse molecular-size axis - a
random message-passing network still encodes atom count, which correlates with
molecular weight - but lacks the organised lipophilicity and ring-count gradients
of the pretrained manifold, consistent with SIGReg having spread the
representation into an isotropic, chemistry-organised distribution.

\subsection{Downstream tasks and evaluation protocol}
\label{sec:eval}

We evaluate the pretrained encoders on two downstream tasks, a niche antibiotic
screen and a standard public benchmark, under the protocols below. Throughout,
the same pretrained checkpoint is compared against the \emph{same} architecture
trained from scratch, so that the pretraining weights are the only difference
between the two arms.

\paragraph{Antibiotic activity.} The Wong et al.\ \cite{Wong2024} dataset
comprises $39{,}312$ compounds assayed against \textit{Staphylococcus aureus}, of
which $512$ ($1.3\%$) are active. We use two splits. The \emph{random} $80/20$
split (stratified by label) reproduces the protocol behind the headline result of
Wong et al.\ \cite{Wong2024}. The \emph{scaffold} split (Bemis-Murcko, with no
scaffold shared between train and test) is the harder, more realistic test of
generalisation to novel chemotypes. Because a scaffold split is only one of many
leakage-free partitions of the same compounds, we do not rely on a single one:
alongside the canonical Bemis-Murcko partition we generate four further
leakage-free scaffold partitions and repeat the entire pretrained-versus-scratch
comparison on each. The metric is the area under the precision--recall curve
(AUPRC), appropriate for the extreme class imbalance. Each model is an ensemble of
five seeds; for the random split, predictions are pooled over five independent
splits ($510$ pooled positives) and reported as the pooled AUPRC with a $95\%$
bootstrap confidence interval. Head-to-head model comparisons use a paired
bootstrap over the pooled test predictions ($10{,}000$ resamples), which removes
the variance shared by two models scored on the same molecules, and we quote
one-sided $p$-values. We compare against the Chemprop result published by Wong
et al.\ \cite{Wong2024} (AUPRC $0.364$ on their random split) and against our own
Chemprop retrained on the identical splits.

\paragraph{\texttt{ogbg-molhiv}.} A single niche assay is a weak basis on which to
judge a pretraining method, so we repeat the scratch-versus-pretrained comparison
on \texttt{ogbg-molhiv} \cite{hu2020open}, a benchmark of ${\sim}41{,}000$
molecules with a fixed Bemis-Murcko scaffold split and ROC-AUC as the metric. We
follow the OGB protocol exactly, selecting the epoch on the validation set and
reporting the corresponding test ROC-AUC averaged over five seeds. Because OGB
does not prescribe the learning rate or the number of frozen-probe epochs, we
sweep both and select by validation, tuning the from-scratch and pretrained models
independently so that neither is disadvantaged.

\paragraph{Frozen probes.} To separate the quality of the learned representation
from what finetuning can recover on its own, we additionally evaluate each encoder
frozen: weights are fixed, each molecule is mean-pooled to its $128$-dimensional
embedding, and a simple head is fit on those features with no gradient ever
reaching the encoder. On the antibiotic scaffold split the head is an
$L^2$-regularised logistic regression, which measures how much signal is encoded
\emph{linearly}. On \texttt{ogbg-molhiv} the head is a random forest, the strong
tabular head that defines the top of the OGB leaderboard, reported as the
mean\,$\pm$\,s.d.\ over ten forest seeds on a fixed encoder. In both cases the
reference points are the same architecture at random initialisation and the
corresponding hand-crafted representation ($217$ RDKit descriptors for the
antibiotic task, a $2048$-bit Morgan fingerprint for \texttt{molhiv}).

\paragraph{Width control and feature-level combination.} A frozen probe that
compares a $128$-dimensional dense embedding against a $2048$-bit sparse
fingerprint confounds the quality of a representation with its dimensionality,
because a random forest with \texttt{max\_features} $=\sqrt{d}$ samples ${\sim}45$
of $2048$ fingerprint columns per split but only ${\sim}11$ of $128$ embedding
columns. We therefore also sweep both representations against \emph{retained
dimensions} under the identical forest: the fingerprint over $64$ to $4096$ bits,
and the embedding truncated two ways - a PCA top-$k$ fitted on the training split
(the best $k$-dimensional linear summary) and a random subset of $k$ raw columns
(no rotation, since a PCA rotation alone is not free on this probe, costing $0.011$
on validation and $0.029$ on test at full width). To
test whether the two representations are complementary we then concatenate them,
sweeping fingerprint width, retained embedding dimensions $k$, and
\texttt{max\_features} (which directly controls how often the embedding block is
sampled at all), fitting the PCA on train only and selecting the reported cell on
the validation split. For cost, these sweeps use smaller forests than the
$2000$-tree, ten-seed probe of Table~\ref{tbl:molhiv-probe} ($500$ trees and five
seeds for the width sweep, $1000$ trees and ten seeds for the concatenation
sweep), which reads a few thousandths low in absolute terms; absolute values are
therefore compared only within a sweep, while head-to-head differences use the
same paired bootstrap over the shared test predictions ($10{,}000$ resamples) as
elsewhere.

\section{Results}

\subsection{Descriptor tree-ensembles match the graph model and exceed the published baseline}

We first establish a strong, simple baseline of tree ensembles over the
$217$ RDKit molecular descriptors, with no graph network and no pretraining.
Table~\ref{tbl:ab-random} reports random-split test AUPRC. A linear model over the
same descriptors reaches only $0.16$, but nonlinear tree ensembles are far
stronger - HistGBM $0.37$, XGBoost $0.43$, and a RandomForest $0.45$ - confirming
that the antibiotic signal carried by the descriptors is strongly nonlinear. The
RandomForest ensemble \textbf{exceeds the published Chemprop result of $0.364$ by
${\sim}0.09$}, using none of the graph machinery.

We compare against a Chemprop model we retrained on the identical splits (pooled
AUPRC $0.407$) with a paired bootstrap over the five pooled splits ($510$
positives), which removes the variance shared by two models scored on the same
molecules. The RandomForest ensemble \emph{significantly} outperforms this
Chemprop ($\Delta\text{AUPRC} = +0.046$, $95\%$ CI $[+0.010, +0.081]$, one-sided
$p = 0.006$); the gradient-boosted XGBoost ensemble sits a step lower, a
statistical tie with Chemprop ($\Delta = +0.025$, $p = 0.086$). We therefore
conclude that \textbf{a descriptor-only tree ensemble matches or exceeds a
graph-based Chemprop and beats the published baseline} - the graph network is not
necessary for this task. The same conclusion holds on the harder scaffold split,
where the RandomForest ensemble reaches AUPRC $0.339$ ($95\%$ CI $[0.269, 0.416]$),
statistically level with the retrained Chemprop $0.312$ (paired $\Delta = +0.023$,
$p = 0.17$) and above the weaker HistGBM $0.281$ and the linear floor $0.112$.
These descriptor baselines set a stringent reference against which any pretraining
gain must be judged.

\begin{table}[t]
  \caption{Antibiotic activity, \emph{random} split, test AUPRC. Every model is
  evaluated by pooling its predictions over the same five random splits ($510$
  pooled positives) and reported as the pooled point AUPRC with a $95\%$ bootstrap
  confidence interval in brackets ($10\,000$ replicates, identical resampling for
  all rows); the published value is quoted from Wong et al.\ \cite{Wong2024} A paired bootstrap
  finds the RandomForest ensemble a significant win over retrained Chemprop
  ($\Delta = +0.046$, $p = 0.006$), XGBoost a tie ($\Delta = +0.025$,
  $p = 0.086$), and GPS+LeJEPA a tie ($\Delta = -0.001$).}
  \label{tbl:ab-random}
  \centering
  \begin{tabular}{lc}
    \toprule
    Model (random split) & Test AUPRC \\
    \midrule
    Logistic regression (descriptors)          & $0.164\ [0.141,\,0.194]$ \\
    HistGBM ensemble (descriptors)             & $0.374\ [0.331,\,0.421]$ \\
    XGBoost ensemble (descriptors)             & $0.432\ [0.386,\,0.478]$ \\
    \textbf{RandomForest ensemble (descriptors)} & $\mathbf{0.453\ [0.407,\,0.500]}$ \\
    \midrule
    Chemprop, retrained (graph)                & $0.407\ [0.361,\,0.457]$ \\
    GPS + LeJEPA (graph)                       & $0.406\ [0.360,\,0.456]$ \\
    \midrule
    Wong et al.~\cite{Wong2024}, published Chemprop & $0.364$ \\
    \bottomrule
  \end{tabular}
\end{table}

\subsection{A frozen probe shows that pretraining learns a better representation}
\label{sec:probe}

We next ask what pretraining does to the representation itself, before any
finetuning is allowed to intervene, by freezing the encoder and fitting a simple
head on the extracted embeddings. On both tasks the pretrained representation is
clearly better than the same architecture at random initialisation.

On the antibiotic scaffold split (Table~\ref{tbl:probe}) the pretrained graph
embedding alone reaches AUPRC $0.159$ under an $L^2$ logistic regression, well
above a randomly-initialised encoder of the same architecture ($0.082$) and above
the $217$ descriptors alone ($0.112$) - direct evidence that LeJEPA \emph{does}
learn antibiotic-relevant structure that random initialisation lacks, and that it
is linearly accessible.

The same pattern holds, more strongly, on \texttt{ogbg-molhiv}. Freezing the
encoder, mean-pooling each molecule to its $128$-dimensional embedding, and
fitting a random forest on top - the strong tabular head that defines the top of
the OGB leaderboard, with no gradient ever reaching the encoder - lifts the LeJEPA
embedding to a test ROC-AUC of $0.788 \pm 0.003$, far above the \emph{same}
architecture at random initialisation ($0.665 \pm 0.004$;
Table~\ref{tbl:molhiv-probe}). This $+0.123$ lift is attributable purely to the
pretrained weights and places the embedding within the $0.75$--$0.79$ band of
published self-supervised GNNs on this benchmark. The same forest applied to a
$2048$-bit Morgan fingerprint scores $0.807 \pm 0.004$ - our faithful reproduction
of the leaderboard's Morgan random-forest entry (ROC-AUC $0.806$) - so the learned
representation is competitive with, though it does not surpass, hand-engineered
circular fingerprints.

\begin{figure}[t]
  \centering
  \includegraphics[width=\linewidth]{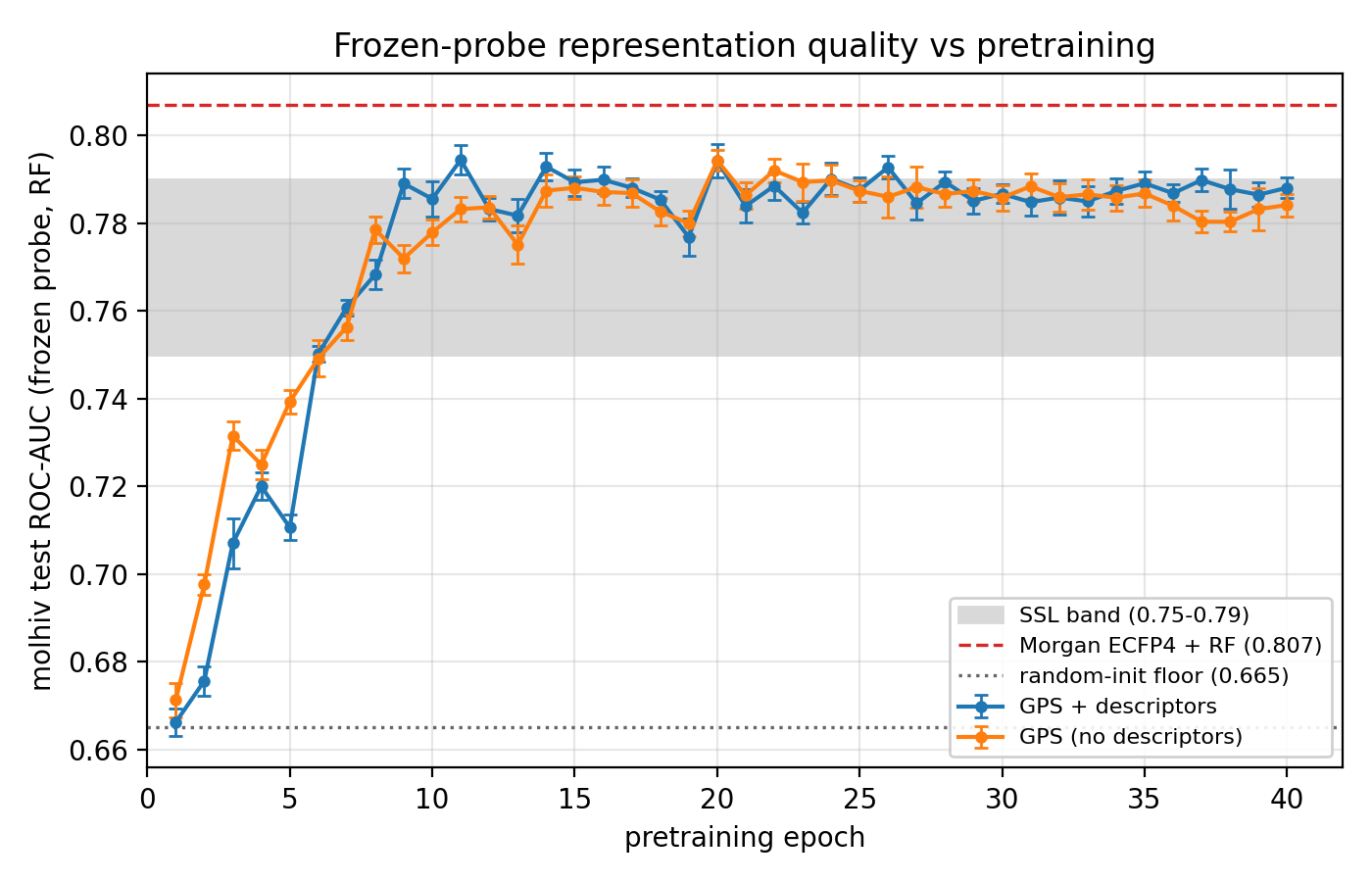}
  \caption{Frozen-probe representation quality across pretraining on
  \texttt{ogbg-molhiv}. At each pretraining epoch the encoder is frozen and a
  random forest (the leaderboard head) is fit on the pooled $128$-dimensional
  embedding; markers are the per-epoch test ROC-AUC and error bars are
  $\pm 1$ s.d.\ over ten forest seeds. Both the descriptor-conditioned run (blue)
  and a descriptor-free run (orange; identical settings, $n_{\text{desc}} = 0$)
  rise off the random-initialisation floor within ${\sim}9$ epochs and then
  plateau just below the Morgan-fingerprint random forest, inside the published
  self-supervised band. The two curves are indistinguishable within seed noise,
  showing that the learned representation forms early and does not depend on
  descriptor conditioning.}
  \label{fig:molhiv-probe-curve}
\end{figure}

Tracking the frozen probe across pretraining checkpoints shows that this
representation forms early and does not depend on descriptor conditioning
(Figure~\ref{fig:molhiv-probe-curve}). The probe ROC-AUC climbs off the
random-initialisation floor within the first ${\sim}9$ epochs and then plateaus
near $0.79$ for the remaining thirty; the $\pm 0.003$ epoch-to-epoch fluctuation on
the plateau is random-forest seed noise, so individual per-epoch maxima carry no
meaning and validation-based checkpoint selection lands anywhere inside the same
plateau. Repeating the entire pretraining run with the descriptor vector removed
($n_{\text{desc}} = 0$, all other settings identical) reproduces this curve almost
exactly, with a plateau-mean test ROC-AUC of $0.785$ versus $0.787$, identical
maxima ($0.794$), and error bars that overlap at every epoch. The descriptors
therefore add no descriptor-specific signal to the pooled graph representation the
probe reads; they serve only the structural role of keeping the global-view
prediction target non-trivial, and the pretraining gain is intrinsic to the graph
encoder.

\begin{table}[t]
  \caption{Frozen linear probe (antibiotic, scaffold split), test AUPRC of an
  $L^2$ logistic regression fit on frozen features. The LeJEPA-pretrained graph
  embedding linearly encodes more antibiotic signal than a random-initialised
  encoder or the descriptors alone. Whether this representational edge survives
  full finetuning is the subject of Section~\ref{sec:finetune}.}
  \label{tbl:probe}
  \centering
  \begin{tabular}{lc}
    \toprule
    Frozen features (linear probe) & Test AUPRC \\
    \midrule
    Random-init graph embedding                 & $0.082$ \\
    RDKit descriptors                           & $0.112$ \\
    \textbf{LeJEPA-pretrained graph embedding}  & $\mathbf{0.159}$ \\
    \bottomrule
  \end{tabular}
\end{table}

\begin{table}[t]
  \caption{\texttt{ogbg-molhiv} frozen probe (scaffold split, test ROC-AUC;
  mean\,$\pm$\,s.d.\ over ten random-forest seeds on a \emph{fixed} encoder, so the
  spread reflects forest-seed variance, not encoder variance). The encoder is
  frozen and a random forest is fit on the pooled embedding, with no gradient
  reaching the encoder. LeJEPA pretraining lifts the embedding far above a
  random-initialised encoder of the same architecture and into the published
  self-supervised band, but does not surpass a Morgan fingerprint fed to the same
  forest.}
  \label{tbl:molhiv-probe}
  \centering
  \begin{tabular}{lc}
    \toprule
    Frozen features + random forest & Test ROC-AUC \\
    \midrule
    Random-init GPS embedding                 & $0.665 \pm 0.004$ \\
    \textbf{GPS + LeJEPA embedding}           & $\mathbf{0.788 \pm 0.003}$ \\
    Morgan fingerprint (ECFP4, $2048$ bits)   & $0.807 \pm 0.004$ \\
    \bottomrule
  \end{tabular}
\end{table}

\subsection{The representational gain does not robustly convert into a finetuning gain}
\label{sec:finetune}

A better frozen representation need not translate into a better finetuned model,
and across four controlled comparisons - two evaluation regimes on the antibiotic
task, a second backbone, and \texttt{ogbg-molhiv} - it largely does not.

\paragraph{Antibiotic, random split: null.} A GPS encoder pretrained with LeJEPA
and finetuned as an ensemble reaches a pooled AUPRC of $0.406$, statistically
indistinguishable from the retrained Chemprop ($0.407$; paired bootstrap
$\Delta = -0.001$) and from the descriptor ensembles (Table~\ref{tbl:ab-random}).
Under independent and identically distributed (i.i.d.)\ evaluation, pretraining
neither helps nor hurts: supervised finetuning on an i.i.d.\ training set recovers
the same signal on its own.

\paragraph{Antibiotic, scaffold split: significant on one partition, not across
five.} On the harder scaffold split, where the test set consists of chemotypes
held out from training, a single canonical Bemis-Murcko partition does show a
gain. Finetuning the same GPS encoder from the LeJEPA checkpoint rather than from
scratch, across five seeds, raises test AUPRC from $0.232 \pm 0.009$ (scratch) to
$0.271 \pm 0.015$ (pretrained) - an improvement in every one of the five seeds -
with a paired bootstrap $\Delta\text{AUPRC} = +0.041$ ($95\%$ CI $[+0.007, +0.079]$,
one-sided $p = 0.010$).

This single-partition result is \emph{not robust}. A scaffold split is only one of
many leakage-free partitions of the same compounds, and a method that genuinely
helps on novel chemotypes should help across them. Repeating the five-seed
pretrained-versus-scratch comparison on four further scaffold partitions
(Table~\ref{tbl:partitions}) shows the effect is strongly partition-dependent. It
is significantly positive on two partitions (the canonical one and partition~4),
non-significant on one, and \emph{negative} on the remaining two, with per-partition
$\Delta\text{AUPRC}$ ranging from $-0.033$ to $+0.054$. Pooling all five partitions
gives $\Delta\text{AUPRC} = +0.013$ ($95\%$ CI $[-0.006, +0.032]$, one-sided
$p = 0.095$), and treating each partition's $\Delta$ as a single observation gives a
mean of $+0.010 \pm 0.038$ ($t = 0.60$); neither reaches significance. The apparent
canonical-split gain is therefore largely an artefact of a favourable partition
rather than a reliable benefit of pretraining.

\paragraph{A second backbone: null.} The same comparison on a Chemprop-style
D-MPNN backbone is null as well. Pretraining leaves it essentially unchanged, with
a five-seed ensemble test AUPRC of $0.277$ (pretrained) versus $0.278$ (scratch) -
$\Delta\text{AUPRC} = -0.000$ ($95\%$ CI $[-0.045, +0.046]$, one-sided $p = 0.48$),
only $2$ of the $5$ seeds improving - so even the fragile, partition-dependent GPS
effect does not carry across architectures.

\paragraph{\texttt{ogbg-molhiv}: null.} Finetuning the pretrained encoder does not
improve on training from scratch on the public benchmark either
(Table~\ref{tbl:molhiv}). The LeJEPA-pretrained encoder reaches a test ROC-AUC of
$0.722 \pm 0.010$ against $0.717 \pm 0.014$ from scratch, a difference of $+0.005$
that lies within one standard deviation. The null is robust to tuning; across the
entire learning-rate by probe-epoch grid the scratch and pretrained configurations
both cluster in the $0.72$--$0.74$ band and overlap completely, with no setting
favouring the pretrained initialisation. Nor does the gap open up in the low-label
regime where pretraining is expected to help most. At the smallest label budgets
($2$--$10\%$ of the training set) the pretrained initialisation is in fact
marginally \emph{worse} than scratch ($\Delta$ ROC-AUC between $-0.03$ and
$-0.07$), because \texttt{molhiv} ROC-AUC already saturates near $0.72$ at $2\%$ of
the data and leaves little headroom for a better initialisation to exploit. The
antibiotic task tells the same low-label story: a few-shot analysis there shows the
frozen pretrained representation does not surpass the descriptors when labels are
scarce.

Taken together, these four comparisons indicate that \textbf{LeJEPA pretraining
yields a measurably better molecular representation - evident under a frozen probe
on two independent tasks (Section~\ref{sec:probe}) - but this advantage does not
robustly convert into a downstream finetuning gain. The apparent scaffold-split
improvement does not survive multi-partition replication, and finetuning is null on
the random split, on \texttt{molhiv}, and with a D-MPNN backbone}.

\begin{table}[t]
  \caption{Antibiotic scaffold split, GPS\,+\,LeJEPA pretrained versus from scratch,
  across five leakage-free Bemis-Murcko scaffold partitions (no scaffold shared
  between train and test within each partition). Each arm is a five-seed ensemble;
  $\Delta\text{AUPRC}$ and the one-sided $p$ come from a paired bootstrap over the
  pooled test predictions ($10\,000$ resamples). The gain is significant on two
  partitions, reversed on two, and the pooled estimate is not significant - the
  single-partition result does not replicate.}
  \label{tbl:partitions}
  \centering
  \begin{tabular}{lcccc}
    \toprule
    Scaffold partition & Scratch & Pretrained & $\Delta$AUPRC & $p$ \\
    \midrule
    canonical    & $0.232$ & $0.271$ & $+0.041$ & $0.010$ \\
    partition 1  & $0.310$ & $0.277$ & $-0.025$ & $0.855$ \\
    partition 2  & $0.313$ & $0.304$ & $+0.015$ & $0.259$ \\
    partition 3  & $0.247$ & $0.223$ & $-0.033$ & $0.911$ \\
    partition 4  & $0.289$ & $0.336$ & $+0.054$ & $0.004$ \\
    \midrule
    \textbf{pooled} & $0.272$ & $0.284$ & $\mathbf{+0.013}$ & $\mathbf{0.095}$ \\
    \bottomrule
  \end{tabular}
\end{table}

\begin{table}[t]
  \caption{\texttt{ogbg-molhiv}, scaffold split, test ROC-AUC (OGB protocol,
  select on validation, report test; mean\,$\pm$\,s.d.\ over five seeds, each at
  its independently validation-selected configuration). LeJEPA pretraining gives
  no significant improvement over training from scratch
  ($\Delta = +0.005$, within one s.d.).}
  \label{tbl:molhiv}
  \centering
  \begin{tabular}{lc}
    \toprule
    Model (\texttt{ogbg-molhiv}, scaffold) & Test ROC-AUC \\
    \midrule
    GPS, from scratch          & $0.717 \pm 0.014$ \\
    GPS + LeJEPA               & $0.722 \pm 0.010$ \\
    \bottomrule
  \end{tabular}
\end{table}

\subsection{The representation is complementary to a fingerprint}
\label{sec:concat}

The finetuning nulls admit two readings. Either the pretrained representation is
genuinely redundant with what supervised learning and hand-crafted features
already supply, or our probes have been comparing it on unfavourable terms. The
second reading deserves a test, because every frozen probe reported so far pits a
$128$-dimensional dense embedding against a $2048$-bit sparse fingerprint under a
random forest, a head that thrives on many weak, decorrelated columns: with
\texttt{max\_features} $=\sqrt{d}$ the forest samples ${\sim}45$ of the $2048$
fingerprint bits at each split but only ${\sim}11$ of the $128$ embedding
dimensions. Dimensionality and representation quality are confounded.

\begin{figure}[t]
  \centering
  \includegraphics[width=\linewidth]{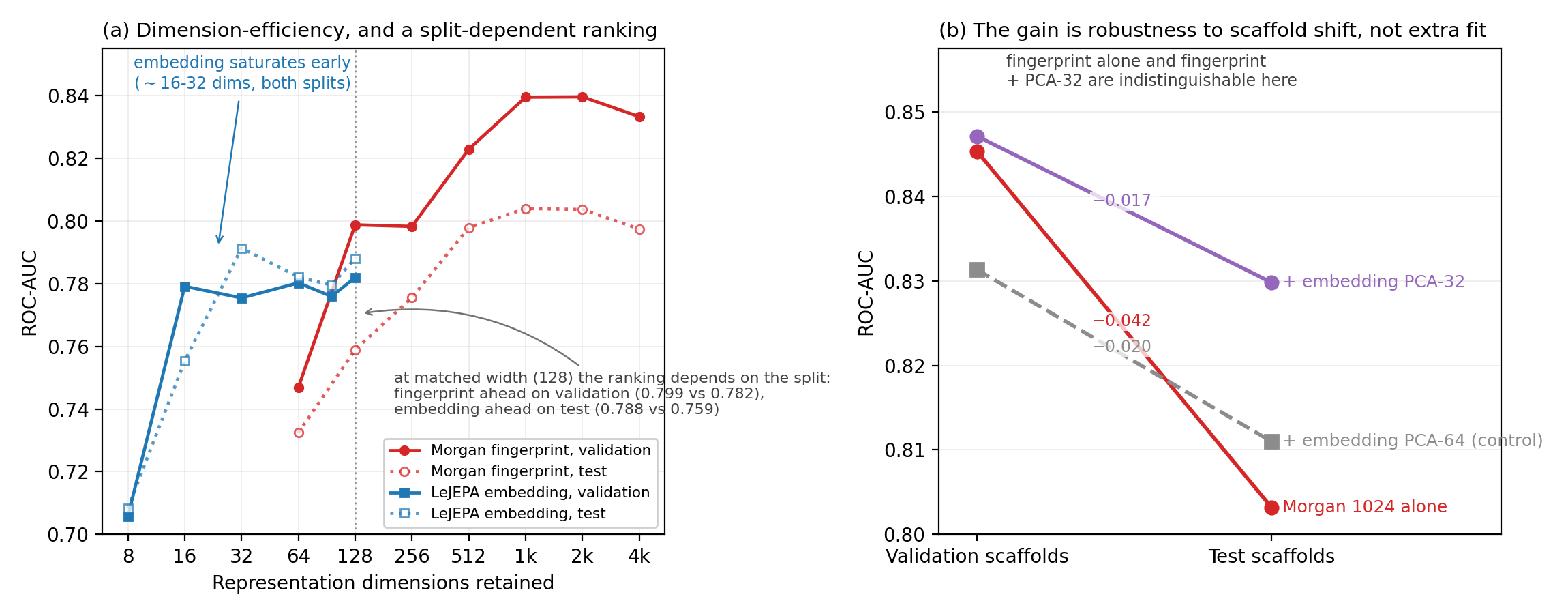}
  \caption{The pretrained embedding is complementary to a fingerprint once its
  informative subspace is isolated. \textbf{(a)} Frozen-probe ROC-AUC against
  retained dimensions under an identical random forest, on the validation (solid) and
  test (dotted) scaffolds. The fingerprint improves to ${\sim}1024$ bits; the
  embedding saturates by ${\sim}16$--$32$ dimensions. At the matched width of $128$
  the ranking inverts between the splits, the fingerprint leading on validation and
  the embedding on test, because the fingerprint loses $0.040$ ROC-AUC across the
  shift while the embedding loses none. \textbf{(b)} Where the
  combination helps. On the validation scaffolds a $1024$-bit fingerprint and the
  same fingerprint plus a $32$-dimensional truncation of the embedding are
  indistinguishable; they separate on the test scaffolds because the combination
  degrades far less across the split ($-0.017$ versus $-0.042$). Restoring the
  diluting dimensions ($k = 64$) removes most of the protection.}
  \label{fig:concat-width}
\end{figure}

\paragraph{The two representations differ in dimension-efficiency.} Sweeping both
against retained width under an identical forest separates them
(Figure~\ref{fig:concat-width}a). On validation the fingerprint improves steadily
with width, from $0.747$ at $64$ bits to $0.839$ at $1024$, where it saturates
($0.840$ at $2048$) and then declines ($0.833$ at $4096$) as its own columns begin to
dilute. The embedding does not improve beyond its first few dimensions: random
subsets of raw columns reach $0.779$ by $16$ dimensions and are flat thereafter
($0.775$, $0.780$, $0.776$, $0.782$ at $32$, $64$, $96$ and $128$), and a PCA
truncation is likewise flat from $16$ onward. The test curves agree for the
fingerprint and for the raw-column embedding. They differ in one place: the PCA
truncation, flat on validation, peaks near $32$ on test and then falls to $0.759$ as
the uninformative tail is restored. The tail is thus not inert but actively
harmful under the shift, which is the same dilution effect the concatenation
experiment below runs into.

The first consequence is that the encoder does not fill the $128$ dimensions it
already has, so representational width is not the binding constraint and a wider
backbone is not indicated; an untrained-backbone control agrees, gaining little from
width ($0.661$, $0.691$, $0.687$ at hidden dimension $128$, $256$, $512$).

The second is more informative, and cuts against the simpler claim that one
representation is better than the other. At \emph{matched} dimensionality the ranking
\emph{inverts between splits}: at $128$ dimensions the fingerprint leads on validation
($0.799$ against $0.782$) and trails on test ($0.759$ against $0.788$). The reason is
that the two degrade very differently across the scaffold shift - the fingerprint
loses $0.040$ ROC-AUC from validation to test, while the embedding loses nothing at
all (it scores $0.006$ \emph{higher} on the shifted split). The fingerprint's
absolute advantage comes entirely from continuing to improve in a width regime where
the embedding has already saturated. So the two representations differ not only in
how much information they extract, but in how much of it survives a change of
chemotype - the same asymmetry the next experiment exploits, visible here without any
combination at all.

\paragraph{Truncated, the embedding adds to the fingerprint.} This also explains why
a naive concatenation fails. Appending all $128$ embedding dimensions to a
$2048$-bit fingerprint yields $2176$ columns of which the forest samples ${\sim}47$
per split, so ${\sim}2.8$ are embedding dimensions and, since only ${\sim}32$ of the
$128$ carry signal, well under one \emph{informative} embedding column is available
at a typical split, while the remaining dimensions supply noise to overfit. Using
each representation near its own optimum instead - the fingerprint at $1024$ bits,
the embedding truncated to a PCA top-$k$ fitted on the training split - changes the
outcome (Table~\ref{tbl:concat}). Selecting the cell with the best \emph{validation}
ROC-AUC, over fingerprint width, $k$ and \texttt{max\_features}, and likewise
selecting the fingerprint-only baseline on validation, gives $1024$ bits plus a
$32$-dimensional truncation, whose test ROC-AUC is $0.830 \pm 0.004$ against the
baseline's $0.803 \pm 0.004$. A paired bootstrap over the shared test predictions
puts the difference at $\Delta = +0.027$ ($95\%$ CI $[+0.003, +0.054]$, one-sided
$p = 0.014$); measured instead against the strongest fingerprint-only cell on test
rather than on validation, it is $+0.025$ ($95\%$ CI $[+0.001, +0.051]$,
$p = 0.019$). Restoring the diluting dimensions removes the effect: at $k = 64$ the
difference falls to $+0.008$ ($95\%$ CI $[-0.021, +0.039]$, $p = 0.29$), a control
that identifies dilution rather than the extra columns as the mechanism.

\paragraph{The gain requires pretraining, not merely extra graph-derived columns.}
A fingerprint gains nothing from a dense block of columns per se, so the comparison
above could in principle reflect the \emph{form} of the added features rather than
what pretraining put in them. Repeating the entire sweep with an untrained backbone
of the same architecture, the same PCA truncation fitted the same way, and the same
selection rule settles this (Table~\ref{tbl:concat}). The untrained embedding is a
far weaker representation on its own ($0.681$ against the pretrained $0.789$), and
appending it to a fingerprint does not help: in the configuration the pretrained
embedding wins on, the untrained one gives $\Delta = -0.003$ ($95\%$ CI
$[-0.029, +0.026]$, $p = 0.59$), and selecting the untrained arm's own best cell on
validation gives $\Delta = -0.013$ ($95\%$ CI $[-0.045, +0.021]$, $p = 0.77$). Across
all twelve cells the pretrained embedding beats its untrained twin in every one, and
the untrained columns are net harmful in most, reaching $-0.030$ at $512$ bits. The
complementarity is therefore a property of what LeJEPA learned, not of the
dimensionality, density or provenance of the added features.

\paragraph{The gain is robustness to scaffold shift, not extra fit.} Where this
improvement lives is more informative than its size, and qualifies it. On the
validation scaffolds the concatenation is not better than the fingerprint alone:
$0.847$ against $0.845$, a difference of $+0.002$ that is well within noise
(Figure~\ref{fig:concat-width}b). The two arms separate only on the test scaffolds,
because they degrade differently across the split: the fingerprint alone loses
$0.042$ ROC-AUC from validation to test, the concatenation loses $0.017$, and the
diluted $k = 64$ control loses an intermediate $0.020$. The embedding is therefore
not adding signal that the validation chemotypes reward; it is adding signal that
survives the change of chemotype. This is consistent with the frozen-probe evidence
that pretraining organises the representation by physico-chemical structure rather
than by task-specific detail, and it is the mechanism one would want from a
pretrained representation. It should nonetheless be read with the caveat it implies:
because the gain is invisible on validation, it cannot be selected \emph{for}, and a
single shifted test set with $130$ actives is a narrow base on which to measure it.

\begin{table}[t]
  \caption{\texttt{ogbg-molhiv} frozen probe, fingerprint and pretrained embedding
  combined (random forest, $1000$ trees, mean\,$\pm$\,s.d.\ over ten forest seeds,
  \texttt{max\_features} $=0.1$). The PCA truncation is fitted on the training split
  only. Configurations are selected on \emph{validation}: doing so picks the
  $1024$-bit fingerprint for the baseline arm and the $1024$-bit fingerprint plus a
  $32$-dimensional truncation for the combined arm (bold). Note that the combined
  representation is \emph{not} better on validation; it is better on the shifted test
  scaffolds, and the $k = 64$ row is the dilution control in which the gain
  disappears. The random-init rows repeat the sweep with an \emph{untrained} backbone
  of the same architecture under the same truncation, and show that graph-derived
  columns as such do not help a fingerprint: only pretrained ones do.}
  \label{tbl:concat}
  \centering
  \begin{tabular}{lccc}
    \toprule
    Frozen features + random forest & dim & Valid ROC-AUC & Test ROC-AUC \\
    \midrule
    \multicolumn{4}{l}{\emph{Each representation alone}} \\
    LeJEPA embedding, PCA-$32$             & $32$   & $0.769$ & $0.789 \pm 0.004$ \\
    Random-init embedding, PCA-$32$        & $32$   & $0.754$ & $0.681 \pm 0.003$ \\
    \midrule
    \multicolumn{4}{l}{\emph{Morgan fingerprint, $1024$ bits, plus a truncated embedding}} \\
    Fingerprint alone                      & $1024$ & $0.845$ & $0.803 \pm 0.004$ \\
    \quad + LeJEPA PCA-$16$                & $1040$ & $0.841$ & $0.827 \pm 0.003$ \\
    \quad \textbf{+ LeJEPA PCA-$32$}       & $1056$ & $\mathbf{0.847}$ & $\mathbf{0.830 \pm 0.004}$ \\
    \quad + LeJEPA PCA-$64$ (dilution control) & $1088$ & $0.831$ & $0.811 \pm 0.003$ \\
    \quad + random-init PCA-$32$ (control) & $1056$ & $0.836$ & $0.800 \pm 0.004$ \\
    \quad + random-init PCA-$64$ (control) & $1088$ & $0.842$ & $0.791 \pm 0.004$ \\
    \midrule
    \multicolumn{4}{l}{\emph{Morgan fingerprint, $512$ bits}} \\
    Fingerprint alone                      & $512$  & $0.821$ & $0.790 \pm 0.005$ \\
    \quad + LeJEPA PCA-$32$                & $544$  & $0.815$ & $0.821 \pm 0.002$ \\
    \quad + random-init PCA-$32$ (control) & $544$  & $0.821$ & $0.773 \pm 0.004$ \\
    \bottomrule
  \end{tabular}
\end{table}

\subsection{Positioning against published baselines}

Two comparisons place these results in context. First, our compact
$128$-dimensional backbone reaches only ${\sim}0.72$ on \texttt{molhiv} even from
scratch, modestly \emph{below} the $0.75$--$0.79$ band occupied by published
self-supervised GNNs on this benchmark (for example a GIN trained from scratch
\cite{xu2019powerful}, MolCLR \cite{wang2022molecular}, and GraphMAE
\cite{hou2022graphmae}). We therefore make no claim to a competitive absolute
number; the informative quantity is the controlled internal
scratch-versus-pretrained comparison, which is null.
Second, the top of the OGB leaderboard \cite{hu2020open} \emph{corroborates} our
antibiotic finding rather than contradicting it. Its strongest \texttt{molhiv}
entry is a molecular-fingerprint plus random-forest ensemble ($\text{ROC-AUC}
\approx 0.85$) that incorporates graph networks only marginally and uses no
pretraining, while the best graph-transformer entry (Graphormer
\cite{ying2021graphormer}, $\approx 0.81$) owes its score to \emph{supervised}
quantum-property transfer rather than self-supervision. As on the antibiotic task,
hand-engineered fingerprints remain the strongest \emph{single} representation; but,
also as on the antibiotic task, LeJEPA pretraining demonstrably learns useful
structure, its frozen embedding far exceeding a random initialisation and reaching
the competitive self-supervised band (Table~\ref{tbl:molhiv-probe}) even though full
finetuning renders that advantage redundant. The strongest representation we obtain
is neither of the two alone but their combination (Section~\ref{sec:concat}), whose
$0.830$ exceeds every single representation we measured and narrows the distance to
that leaderboard-topping fingerprint ensemble - achieved by adding $32$ columns to
$1024$, rather than by scaling either component.

\section{Discussion and conclusion}
\label{sec:conclusion}

We adapted LeJEPA, a predictor-free joint-embedding predictive architecture
regularised by Sketched Isotropic Gaussian Regularisation (SIGReg), to molecular
graphs, and evaluated whether the resulting self-supervised pretraining improves
downstream property prediction. Our central finding is that the benefit of
pretraining is real at the level of the learned \emph{representation} - a frozen
probe on the pretrained embedding clearly beats a randomly-initialised encoder on
two independent tasks - but does not robustly carry through to a downstream
\emph{finetuning} gain, the one apparently-significant transfer result failing to
replicate across scaffold partitions. That benefit is not, however, unusable: it
can be realised at the feature level, by isolating the embedding's informative
subspace and combining it with a fingerprint.

Four results support this conclusion. First, on the Wong et al.\ \cite{Wong2024} antibiotic
dataset a tree ensemble over $217$ RDKit descriptors - using no graph
network and no pretraining - matches or beats a graph-based Chemprop and exceeds the
published baseline, establishing that the graph model is not required for this
task and setting a stringent reference for any pretraining gain. Second, LeJEPA
pretraining nonetheless learns a representation that, frozen, linearly encodes more
antibiotic-relevant signal than a randomly-initialised encoder ($0.159$ vs $0.082$
AUPRC) and that on \texttt{ogbg-molhiv} reaches the published self-supervised band
($0.788$ vs $0.665$ ROC-AUC under a frozen random-forest probe, a $+0.123$ lift) -
a real and reproducible representational improvement. Third, this representational
advantage does not robustly convert into a downstream finetuning gain. The one
finetuning result that reaches significance - the antibiotic scaffold split, $\Delta
\text{AUPRC} = +0.041$ on a single canonical partition - does not replicate across
five scaffold partitions, where it ranges from $-0.033$ to $+0.054$ and the pooled
estimate is $+0.013$ ($95\%$ CI $[-0.006, +0.032]$, $p = 0.095$, not significant).
Finetuning is likewise null under in-distribution (random-split) evaluation, on
which supervised learning recovers an equivalent solution unaided, and on
\texttt{ogbg-molhiv}. The effect is also architecture-dependent. The same
pretraining gives a Chemprop-style D-MPNN backbone no measurable benefit even on the
canonical scaffold split ($\Delta\text{AUPRC} = -0.000$, $95\%$ CI $[-0.045,
+0.046]$, $p = 0.48$).

Fourth, and against that run of nulls, the representational advantage \emph{does}
convert downstream when it is combined with a fingerprint instead of being finetuned.
A width-controlled analysis shows the embedding saturates at ${\sim}16$--$32$
effective dimensions, and that at matched dimensionality the ranking inverts between
splits: the fingerprint leads on validation and the embedding on the shifted test
scaffolds, which it crosses with essentially no loss against the fingerprint's
$0.040$. Truncating the embedding to that subspace
and concatenating it with a $1024$-bit fingerprint raises \texttt{ogbg-molhiv} test
ROC-AUC from $0.803$ to $0.830$ ($\Delta = +0.027$, $95\%$ CI $[+0.003, +0.054]$,
$p = 0.014$), with a $k = 64$ dilution control in which the gain disappears. An
untrained backbone put through the identical pipeline gains nothing ($\Delta =
-0.003$, $p = 0.59$), so this is a consequence of pretraining rather than of adding
dense graph-derived columns to a fingerprint.

The coherent reading is that LeJEPA pretraining supplies representational
structure that is most useful precisely when the downstream task demands
generalisation beyond the training distribution, and is otherwise redundant with
what supervised learning and hand-engineered descriptors already provide. The
concatenation result makes this concrete rather than merely interpretive: the
combined representation is no better than the fingerprint on the validation
scaffolds ($0.847$ vs $0.845$) and better only on the test scaffolds, because it
loses far less across the shift ($-0.017$ vs $-0.042$). What pretraining contributes
is robustness to a change of chemotype, which is invisible to any metric computed
on data drawn like the training set. This is consistent with the broader
observation, reinforced by the descriptor and fingerprint baselines on both tasks,
that low-order physico-chemical features carry much of the signal in these
property-prediction problems.

Several limitations bound these conclusions. The downstream finetuning evidence is
the weak point of the pretraining case. The one significant transfer result, on the
canonical antibiotic scaffold partition, does not survive replication across five
scaffold partitions (Table~\ref{tbl:partitions}), so we report it as
partition-dependent rather than as a reliable gain. The robust, reproducible signal
is at the level of the representation (the frozen probes on both tasks), and why
finetuning fails to exploit a representation that a random forest can exploit
remains the central open problem this work leaves. The concatenation result carries
its own caveats. It rests on one benchmark, one split and $130$ test actives; the
truncation dimension is a hyperparameter, chosen on validation, which is defensible
but must be described as fitted; and because the benefit appears only under the
scaffold shift and not on validation, it is a quantity that cannot be selected for
in the ordinary way. Our compact encoder attains absolute
scores below the strongest published self-supervised graph networks, so our claims
concern the controlled scratch-versus-pretrained contrast rather than leaderboard
standing. Finally, the pretraining corpus does not fully cover the downstream
chemical space, which upper-bounds the achievable transfer. As to where effort is
best directed, our width analysis argues \emph{against} simply enlarging the
encoder, since the present one does not fill the dimensions it has; the more
promising directions are pretraining objectives or auxiliary signals that are
complementary to - rather than redundant with - molecular descriptors, evaluation
designed around distribution shift rather than in-distribution accuracy, and
treating the pretrained embedding as a component to be combined with established
representations rather than as a replacement for them.

\section*{Declaration of generative AI use}

The authors used a generative AI assistant (Anthropic's Claude, through the
Claude Code command-line interface) throughout this work. Its use covered
implementation and analysis - writing and refactoring parts of the pretraining,
finetuning, evaluation and plotting code, and helping to build the statistical
analysis scripts - and the preparation of this manuscript, where it assisted with
drafting, editing and \LaTeX{} formatting. It was not used to generate
experimental data. All research questions, experimental designs and
interpretations are the authors' own; every training and evaluation run reported
here was launched and supervised by the authors, and every number, table and
figure was checked against the underlying experimental output. The authors
reviewed and edited all AI-assisted code and text, and take full responsibility
for the content of this article.

\printbibliography

\end{document}